\documentclass[12pt]{article}
\def\int {\intop \limits}

\def\fnote#1{\footnote}
\begin{document}

\renewcommand \theequation{\thesection.\arabic{equation}}

\title{Spectrum and polarization of coherent and incoherent radiation and
the LPM effect in oriented single crystal}
\author{V. N. Baier
and V. M. Katkov\\
Budker Institute of Nuclear Physics,\\ Novosibirsk, 630090, Russia}

\maketitle

\begin{abstract}
The spectrum and the circular polarization of radiation from
longitudinally polarized high-energy electrons in oriented single
crystal are considered using the method which permits inseparable
consideration of both the coherent and the incoherent mechanisms of
photon emission. The spectral and polarization properties of
radiation are obtained and analyzed. It is found that in some part
of spectral distribution the influence of multiple scattering (the
Landau-Pomeranchuk-Migdal (LPM) effect) attains the order of 7
percent. The same is true for the influence of multiple scattering
on the polarization part of the radiation intensity. The degree of
circular polarization of total intensity of radiation is found. It
is shown that the influence of multiple scattering on the photon
polarization is similar to the influence of the LPM effect on the
total intensity of radiation: it appears only for relatively low
energies of radiating electron and has the order of 1 percent, while
at higher energies the crystal field action excludes the LPM effect.

\end{abstract}

\newpage

\section{Introduction}

The study of processes with participation of polarized electrons and
photons permits to obtain the important physical information.
Because of this reason the experiments with use of polarized
particles are performed and are planning in many laboratories (CERN,
Jefferson Nat Accl Fac, SLAC, BINP, etc). In this paper the
polarization effects are considered in the frame of general theory
developed by authors \cite{BK1}, which includes both the coherent
and the incoherent mechanisms of radiation from high-energy
electrons in an oriented single crystal. The influence of multiple
scattering on the radiation process including polarization effects
is analyzed. The study of radiation in oriented crystals is
continuing and new experiments are performed recently see
\cite{KMU}, \cite{BKi}.

The general expression for the energy loss of the longitudinally
polarized electron in oriented crystal was found in \cite{BK} (see
Eq.(2.7))
\begin{eqnarray}
&& dE_{\xi} = -\frac{\alpha m^2}{8
\pi^2}\frac{d^3k}{\varepsilon\varepsilon'} \int
\frac{d^3r}{V}F(\textbf{r}, \vartheta_0) \int e^{-iA}
\left[\varphi_1(\xi)+\frac{1}{4}\varphi_2(\xi)\gamma^2\left(\textbf{v}_1-
\textbf{v}_2\right)^2\right]dt_1dt_2,
\nonumber \\
&& A=\frac{\omega \varepsilon}{2\varepsilon'}\int_{t_1}^{t_2}
\left[\frac{1}{\gamma^2}+(\textbf{n}-\textbf{v}(t))^2\right]dt,
\nonumber \\
&&\varphi_1(\xi)=1+\xi\frac{\omega}{\varepsilon}, \quad
\varphi_2(\xi)=(1+\xi)\frac{\varepsilon}{\varepsilon'}+
(1-\xi)\frac{\varepsilon'}{\varepsilon}. \label{1}
\end{eqnarray}
where $dE_{\xi} = \omega dw_{\xi}$, $dw_{\xi}$ is the probability of
radiation, see e.g. Eq.(4.2) in \cite{BKS}, $\omega$ and
$\varepsilon$ are the photon and electron energy,
$\alpha=e^2=1/137$, the vector $\textbf{k}$ is the photon momentum,
 ${\bf n}={\bf k}/|{\bf k}|$,
$\xi=\lambda\zeta,~\lambda=\pm 1$ is the helicity of emitted photon,
$\zeta =\pm 1$ is the helicity of the initial electron,
$F(\textbf{r}, \vartheta_0)$ is the distribution function of
electron in the transverse phase space depending on the angle of
incidence $\vartheta_0$ of the electron on crystal,
$\textbf{v}_1=\textbf{v}(t_1)$ is the electron velocity (see
\cite{BKS}, Sec.16.2).

The degree of the circular polarization of radiation is defined by
Stoke's parameter $\xi_{2}$:
\begin{equation}
\xi_{2}=\Lambda (\mbox{\boldmath$\zeta$}{\bf v}),\quad \Lambda =
\frac{dE_+ - dE_-}{dE_+ + dE_-}, \label{1a}
\end{equation}
where the quantity $(\mbox{\boldmath$\zeta$}{\bf v})$ defines the
longitudinal polarization of the initial electrons, $dE_+$ and
$dE_-$ is the energy loss for $\xi$=+1 and $\xi$=-1 correspondingly.

It should be noted that a few different spin correlations are known
in an external field. But after averaging over directions of crystal
field only the considered here longitudinal polarization survives.

In \cite{BK} the polarization effects in the coherent radiation
which dominates at high electron energies ($\varepsilon \gg 1$~GeV
for main axes of heavy elements, e.g. tungsten crystal) was studied.
At intermediate energies the incoherent radiation contributes
essentially and the contributions of both mechanisms should be taken
into account. Recently authors developed the method which permits
indivisible consideration of both the coherent and the incoherent
mechanisms of photon emission in oriented crystals \cite{BK1}.

Basing on Eqs.(18) and (19) of \cite{BK1} (see also Eqs. (7.89) and
(7.90) in \cite{BKS}) and using Eq.(\ref{1}) one can obtain the
general expression for the intensity of radiation from
longitudinally polarized electrons which includes the coherent and
incoherent contributions and the Landau-Pomeranchuk-Migdal (LPM)
effect:
\begin{eqnarray}
&& dI_{\xi}(\varepsilon,y)=dI_0(\varepsilon,y)+\xi
dI_1(\varepsilon,y)=\frac{\alpha m^2}{2\pi} \frac{y dy}{1-y}
\int_0^{x_0}\frac{dx}{x_0}G_{r\xi}(x, y),
\nonumber \\
&&G_{r\xi}(x, y)=\int_0^{\infty} F_{r\xi}(x, y, t)dt
-r_{3\xi}\frac{\pi}{4},
\nonumber \\
&& F_{r\xi}(x, y, t)={\rm Im}\left\lbrace
e^{\varphi_1(t)}\left[r_{2\xi}\nu_0^2
(1+ib_r)\varphi_2(t)+r_{3\xi}\varphi_3(t) \right]
\right\rbrace,\quad b_r=\frac{4\chi^2(x)}{u^2\nu_0^2},
\nonumber \\
&& y=\frac{\omega}{\varepsilon}, \quad u=\frac{y}{1-y},\quad
\varphi_1(t)=(i-1)t+b_r(1+i)(\varphi_2(t)-t),
\nonumber \\
&&
\varphi_2(t)=\frac{\sqrt{2}}{\nu_0}\tanh\frac{\nu_0t}{\sqrt{2}},\quad
\varphi_3(t)=\frac{\sqrt{2}\nu_0}{\sinh(\sqrt{2}\nu_0t)}, \label{2}
\end{eqnarray}
where
\begin{eqnarray}
&&r_{2\xi}=\frac{1}{2}\left(r_2+\xi r_{21}\right),\quad
r_2=1+(1-y)^2,\quad r_{21}=2y-y^2
\nonumber \\
&&r_{3\xi}=\frac{1}{2}\left(r_3+\xi r_{31}\right),\quad
r_3=2(1-y),\quad r_{31}=2y(1-y),
\nonumber \\
&&\nu_0^2=\frac{1-y}{y} \frac{\varepsilon}{\varepsilon_c(x)},
\label{3}
\end{eqnarray}
The intensity for unpolarized electrons $dI_0(\varepsilon,y)$ was
obtained in \cite{BK1}, the polarization term $dI_1(\varepsilon,y)$
is found here.

The situation is considered when the electron angle of incidence
$\vartheta_0$ (the angle between electron momentum {\bf p} and the
axis (or plane)) is small $\vartheta_0 \ll V_0/m$. The axis
potential (see Eq.(9.13) in \cite{BKS}) is taken in the form
\begin{equation}
U(x)=V_0\left[\ln\left(1+\frac{1}{x+\eta} \right)-
\ln\left(1+\frac{1}{x_0+\eta} \right) \right], \label{4}
\end{equation}
where
\begin{equation}
x_0=\frac{1}{\pi d n_a a_s^2}, \quad  \eta_1=\frac{2
u_1^2}{a_s^2},\quad x=\frac{\varrho^2}{a_s^2}, \label{5}
\end{equation}
Here $\varrho$ is the distance from axis, $u_1$ is the amplitude of
thermal vibration, $d$ is the mean distance between atoms forming
the axis, $a_s$ is the effective screening radius of the potential.
The parameters in Eq.(\ref{4}) were determined by means of fitting
procedure.

The local value of parameter $\chi(x)$  which determines the
radiation probability in the field Eq.(\ref{4}) is
\begin{equation}
\chi(x)=-\frac{dU(\varrho)}{d\varrho}\frac{\varepsilon}{m^3}=\chi_s
\frac{2\sqrt{x}}{(x+\eta)(x+\eta+1)},\quad \chi_s=\frac{V_0
\varepsilon}{m^3a_s}\equiv \frac{\varepsilon}{\varepsilon_s}.
 \label{5}
\end{equation}
For an axial orientation of crystal the ratio of the atom density
$n(\varrho)$ in the vicinity of an axis to the mean atom density
$n_a$ is (see \cite{BK1})
\begin{equation}
\frac{n(x)}{n_a}=\xi(x)=\frac{x_0}{\eta_1}e^{-x/\eta_1},\quad
\varepsilon_0=\frac{\varepsilon_e}{\xi(0)}, \quad
\varepsilon_e=\frac{m}{16\pi Z^2\alpha^2\lambda_c^3n_aL_0}.\label{6}
\end{equation}

The functions and values in Eqs.(\ref{2}) and (\ref{3}) are
\begin{eqnarray}
&&\varepsilon_c(x)=
\frac{\varepsilon_e(n_a)}{\xi(x)g(x)}=\frac{\varepsilon_0}{g(x)}e^{x/\eta_1},\quad
L_0=\ln(ma)+ \frac{1}{2}-f(Z\alpha),
\nonumber \\
&&g(x)=g_0+\frac{1}{6 L_0}\left[\ln
\left(1+\frac{\chi^2(x)}{u^2}\right)+\frac{6 D_{sc}\chi^2(x)}
{12u^2+\chi^2(x)}\right],
\nonumber \\
&&
g_0=1+\frac{1}{L_0}\left[\frac{1}{18}-h\left(\frac{u_1^2}{a^2}\right)\right],\quad
a=\frac{111Z^{-1/3}}{m}, \quad f(\xi)=\sum_{n=1}^{\infty}
\frac{\xi^2}{n(n^2+\xi^2)},
\nonumber \\
&& h(z)=-\frac{1}{2}\left[1+(1+z)e^{z}{\rm Ei}(-z) \right],\label{7}
\end{eqnarray}
where the function $g(x)$ determines the effective logarithm using
the interpolation procedure:$L=L_0g(x)$, see Eq.(14) in \cite{BK1},
$D_{sc}=2,3008$ is the constant entering in the radiation spectrum
at $\chi/u \gg 1$, see Eq.(7.107) in \cite{BKS},~ Ei($z$) is the
integral exponential function, $f(\xi)$ is the Coulomb correction.

It follows from Eqs.(\ref{1a}) and (\ref{2}) that the circular
polarization of radiation is
\begin{equation}
\xi_{2}=
\frac{dI_1(\varepsilon,y)}{dI_0(\varepsilon,y)}(\mbox{\boldmath$\zeta$}{\bf
v}), \label{8}
\end{equation}

\section{The spectral distribution of radiation}
\setcounter{equation}{0}

The expression for $dI_{\xi}$ Eq.(\ref{2}) includes both the
coherent and incoherent contributions as well as the influence of
the multiple scattering (the LPM effect) on the photon emission
process.

The probability of the coherent radiation
$dI_0^{coh}(\varepsilon,y)$ is the first term ($\nu_0^2=0$) of the
decomposition of Eq.(\ref{2}) over $\nu_0^2$. This probability is
contained in Eq.(17.7) of \cite{BKS}. The polarization term in the
probability of the coherent radiation $dI_1^{coh}(\varepsilon,y)$ is
the first term of the decomposition of $dI_1$ in Eq.(\ref{2}) over
$\nu_0^2$. The expression $dI_0^{coh}+\xi dI_1^{coh}$ coincides with
the term containing $R_0(\lambda)$ in Eq.(3.5) of \cite{BK}.

The intensity of the incoherent radiation
$dI_{0}^{inc}(\varepsilon,y)$ is the second term ($\propto \nu_0^2$)
of the mentioned decomposition of $dI(\varepsilon,y)$ \cite{BK1}.
The expression for $dI_{0}^{inc}(\varepsilon,y)$ follows also from
Eq.(21.21) in \cite{BKS}). The polarization term
$dI_{1}^{inc}(\varepsilon,y)$ is correspondingly the second term
($\propto \nu_0^2$) of decomposition of $dI_1(\varepsilon,y)$:
\begin{equation}
dI_{0,1}^{inc}(\varepsilon,y)=\frac{\alpha m^2}{60\pi}
\frac{\varepsilon}{\varepsilon_0}\int_0^{x_0}g(x)e^{-x/\eta_1}
dJ_{0,1}^{inc}(\chi,y)\frac{dx}{x_0}, \label{11}
\end{equation}
here $\chi=\chi(x)$, the notations is given in Eqs.(\ref{5}),
(\ref{6}) and (\ref{7}), $dJ_{0,1}^{inc}(\chi)$ can be written as
\begin{eqnarray}
&&dJ_{0}^{inc}(\chi,y)=\left[y^2(f_1(z)+f_2(z))+2(1-y)f_2(z)\right]dy,
\nonumber \\
&& dJ_{1}^{inc}(\chi,y)=\left[y^2(f_1(z)-f_2(z))+2yf_2(z)\right]dy,
\nonumber \\
&&z=\left(\frac{y}{\chi(1-y)}\right)^{2/3}, \label{12}
\end{eqnarray}
the functions $f_1(z)$ and $f_2(z)$ are defined in the just
mentioned equation in \cite{BKS}:
\begin{eqnarray}
&& f_1(z)=z^4\Upsilon(z)-3z^2\Upsilon'(z)-z^3,
\nonumber \\
&& f_2(z)=(z^4+3z)\Upsilon(z)-5z^2\Upsilon'(z)-z^3, \label{13}
\end{eqnarray}
here $\Upsilon(z)$ is the Hardy function:
\begin{equation}
\Upsilon(z)=\int_0^{\infty}\sin\left(z\tau+\frac{\tau^3}{3}\right)d\tau.
\label{14}
\end{equation}

For intermediate energies, where both the coherent and the
incoherent contributions to the total intensity of radiation are
essential, the spectral distribution of intensity
$dI_{0}(\varepsilon,y)$ is shown in Fig.1. The calculation was done
for axis $<111>$ of tungsten at low temperature T=100 K (parameters
of crystal are given Table 1). These spectra describe radiation in
thin targets when one can neglect the energy loss of projectile. It
is seen that the phenomena under consideration become apparent at
relatively low energy. For $\varepsilon=0.3$~GeV, $dI_0^{coh} \simeq
dI_0^{inc}$ at $y \simeq 0.1~(\omega \simeq 60~MeV)$ while for lower
photon energy the coherent contribution dominates and for higher
photon energy the incoherent contribution dominates. For
$\varepsilon=1$~GeV, $dI_0^{coh} \simeq dI_0^{inc}$ at $y \simeq
0.28~(\omega \simeq 280~MeV)$ and for $\varepsilon=3$~GeV,
$dI_0^{coh} \simeq dI_0^{inc}$ at $y \simeq 0.54~(\omega \simeq
1.6~GeV)$. All spectrum curves have very steep (exponential) right
slope the location of which is defined by the electron energy.

The next terms of decomposition of the intensity $dI_0(\varepsilon,
y)$ over $\nu_0^2$ describe  the influence of multiple scattering on
the radiation process, the LPM effect. The different contributions
to that part of the spectrum, where the coherent and the incoherent
contributions are comparable, are shown in Fig.2. The difference
shown by curve 5 arises due to the LPM effect. We define the
contribution of the LPM effect into spectral distribution, by
analogy with \cite{BK1}, as
\begin{equation}
\Delta_s=-\frac{dI_0-dI_0^{coh}-dI_0^{inc}}{dI_0} \label{s15}
\end{equation}
The function $\Delta_s(y)$ is shown in Fig.3. The curve 1 for
$\varepsilon=0.3~$ GeV reaches the maximum 6.64 \% at y=0.18, the
curve 2 for $\varepsilon=1~$ GeV reaches the maximum 6.87 \% at
y=0.44 and the curve 3 for $\varepsilon=3~$ GeV reaches the maximum
7.32 \% at y=0.7.

At room temperature (T=293 K) for axis $<111>$ in tungsten for the
electron energy $\varepsilon=10~$ GeV the different contributions to
that part of the spectrum where the coherent and the incoherent
contributions are comparable are shown in Fig.4. In this case the
maximum of the function $\Delta_s(y) \simeq 6.03$ \% is reached at
y=0.82.

All the curves in Fig.2 have nearly the same height of the maximum
and the position of the maximum is defined roughly by the expression
$u_m \simeq 3\varepsilon/\varepsilon_0~(u\equiv y/(1-y))$. Such
scaling in terms of $u$ is the consequence of the following
representation of the spectral inverse radiation length (the
intensity spectrum Eq.(\ref{2}))
\begin{equation}
\frac{dL_{rad}^{-1}}{dy}=\frac{1}{\varepsilon}\frac{dI_0(\varepsilon,
y)}{dy}= r_2(y)R_2\left(\frac{\varepsilon}{u}\right)+
r_3(y)R_3\left(\frac{\varepsilon}{u}\right)\label{a15}
\end{equation}
In the high energy limit $\varepsilon \gg \varepsilon_0$ the maximum
of the LPM effect is situated at the very end of the spectrum. In
this limit $r_2\simeq 1-O(\varepsilon_0/ \varepsilon)$ and
$r_3\simeq O(\varepsilon_0/ \varepsilon)$ and the scaling
(dependence on the combination $\varepsilon/u$ only) of each of the
two $R_{2,3}$ terms gets over into scaling of the whole expression
for the spectral inverse radiation length.

In the maximum of the LPM effect the coherent contribution into
spectral radiation intensity is relatively small: less than 10\%.
Therefore the right slope of curves in Fig.3 is described by
formulas of the LPM effect in a medium (incoherent radiation) with
corrections due to action of the crystal field. Far of the maximum
at $u \gg \varepsilon/ \varepsilon_0$ one has $R_{2,3}^{coh}=0$ and
the terms $\propto \nu_0^6$ in the decomposition of the functions
$R_{2,3}$ (which includes the crystal field corrections) have the
form
\begin{eqnarray}
&& \Lambda_2=\frac{\varepsilon^2g_0^2}{3\varepsilon_0^2u^2}\left(1+
377\frac{\overline{\chi^2}}{u^2}\right),\quad
\Lambda_3=-\frac{\varepsilon^2g_0^2}{3\varepsilon_0^2u^2}\left(\frac{31}{63}+
\frac{2704}{15}\frac{\overline{\chi^2}}{u^2}\right)
\nonumber \\
&& \Lambda_s=\frac{R_s^{inc}-R_s}{R_s^{inc}(\chi=0)},\quad
\overline{\chi^2}=\int_0^{\infty}\chi^2(x)e^{-3x/\eta_1}\frac{dx}{\eta_1}
\label{b15}
\end{eqnarray}
Here the terms independent on field coincide with corresponding
terms in Eq.(3.6) in \cite{BK3}, the corrections depending on
crystal field are calculated in this paper. At the left slope of the
curves in Fig.3 the coherent contribution dominates (see Fig.1), the
relative contribution of incoherent radiation diminishes and the LPM
effect is only its small part.

Degree of the circular polarization of radiation Eq.(\ref{8}) is
shown in Fig.5. The curves for energies $\varepsilon=0.3, 1, 3~$GeV
coincide with each other inside thickness of line. For any mechanism
of radiation $\xi_{2} \simeq y (\mbox{\boldmath$\zeta$}{\bf v})$ for
$y \ll 1$ (see Eq.(2.9)) in \cite{BK} and $\xi_{2} \rightarrow 1$
for $y \rightarrow 1$ as a consequence of the helicity transfer from
an electron to photon.

The next terms of decomposition of the intensity
$dI_1(\varepsilon,y)$ over $\nu_0^2$ describe  the influence of
multiple scattering on the polarization part of the spectral
intensity of radiation. We define the contribution of this effect
into the polarization part as
\begin{equation}
\Delta_{s1}=-\frac{dI_1-dI_1^{coh}-dI_1^{inc}}{dI_1} \label{s16}
\end{equation}
The function $\Delta_{s1}(y)$  for $\varepsilon=0.3~$ GeV reaches
the maximum 6.78 \% at y=0.18, for $\varepsilon=1~$ GeV reaches the
maximum 7.09 \% at y=0.44 and for $\varepsilon=3~$ GeV reaches the
maximum 7.43 \% at y=0.7. It is seen that the maximum positions are
situated at the same photon energy as in $\Delta_{s}(y)$ (see Fig.3)
and their values are very close to these values in the unpolarized
part. All this means that the multiple scattering is affecting
similarly on the unpolarized spectrum described by
$dI_0(\varepsilon, y)$ and the polarization term described by
$dI_1(\varepsilon, y)$.

The influence of the multiple scattering on the photon polarization
degree may be also characterized by
\begin{equation}
\Delta_{s\xi}=-\frac{\xi_{s2}^T-\xi_{s2}^{ci}}{\xi_{s2}^{ci}}
=\Delta_{s1}-\Delta_s,\quad
\xi_{s2}^T=\frac{dI_1(\varepsilon,y)}{dI_0(\varepsilon,y)},\quad
\xi_{s2}^{ci}=\frac{dI_1^{coh}+dI_1^{inc}}{dI_0^{coh}+dI_0^{inc}},\
 \label{a16}
\end{equation}
Since value $\Delta_{s1}$ is very close to $\Delta_{s}$ the value of
$\Delta_{s\xi}$ is much smaller than both $\Delta_{s}$ and
$\Delta_{s1}$.

\section{Effect for the total intensity of radiation}
\setcounter{equation}{0}

Now we turn to analysis of the polarization effects for the total
intensities of radiation
\begin{equation}
I_{\xi}(\varepsilon)=I_0(\varepsilon)+\xi I_1(\varepsilon),\quad
I_0(\varepsilon)=\int_{y=0}^{y=1}dI_0(\varepsilon),\quad
I_1(\varepsilon)=\int_{y=0}^{y=1} dI_1(\varepsilon).
 \label{9}
\end{equation}
The integral degree of the circular polarization of the radiation
intensity in a crystal is given by the ratio
$\xi_2^T=I_{1}(\varepsilon)/I_0(\varepsilon)$.

In \cite{BK1} it was shown that the total intensity $I(\varepsilon)$
contains both the coherent and incoherent contributions as well as
the influence of the multiple scattering (the LPM effect) on the
process under consideration. The same is true for the polarization
term $I_1(\varepsilon)$. The intensity of coherent radiation
$I^F(\varepsilon)\equiv I_0^{coh}(\varepsilon)$ is the first term
($\nu_0^2=0$) of the decomposition of $I(\varepsilon)$ over
$\nu_0^2$. Its explicit representation is given by Eqs.(25) and (26)
in \cite{BK1}. The coherent polarization term
$I_1^{coh}(\varepsilon)$ is the first term ($\nu_0^2=0$) of the
decomposition of $I_1(\varepsilon)$ over $\nu_0^2$. Both can be
written in the form
\begin{eqnarray}
&&
I_{0,1}^{coh}(\varepsilon)=\int_{0}^{x_0}J_{0,1}^{coh}(\chi)\frac{dx}{x_0},
\nonumber \\
&& J_{0,1}^{coh}(\chi)=i\frac{\alpha m^2}{2\pi}
\int_{\lambda-i\infty}^{\lambda+i\infty}
\left(\frac{\chi^2}{3}\right)^s \Gamma\left(1-s\right)
\Gamma\left(3s-1\right)(2s-1)a_{0,1}\frac{ds}{\cos\pi s},
\nonumber \\
&&a_0=s^2-s+2,\quad a_1=\frac{11}{6}(1-s),\quad
\frac{1}{3}<\lambda<1.\label{10}
\end{eqnarray}
where $J_0^{coh}(\chi)$ is the radiation intensity  and
$J_1^{coh}(\chi)$ is the contribution of the circular polarization
of radiation in external field (see Eqs.(4.50), (4.51) and (4.84) in
\cite{BKS}). The representation (\ref{10}) is convenient both for
the analytical and numerical calculation. The degree of circular
polarization of the coherent radiation in a crystal we define by the
ratio $\xi_2^{coh}=I_{1}^{coh}/I_{0}^{coh}$.

In \cite{BK1} the new representation of the function
$J_{0}^{inc}(\chi)$ was obtained, which is suitable for both
analytical and numerical calculation. The same procedure can be
applied to $J_{1}^{inc}(\chi)$. As a result we get
\begin{equation}
J_{0,1}^{inc}(\chi)=\frac{i\pi}{2}\int_{\lambda-i\infty}^{\lambda+i\infty}
\frac{\chi^{2s}}{3^s}
\frac{\Gamma(1+3s)}{\Gamma(s)}R_{0,1}(s)\frac{ds}{\sin^2\pi s},\quad
-\frac{1}{3} < \lambda <0 \label{15}
\end{equation}
where
\begin{equation}
R_0(s)=15+43s+31s^2+28s^3+12s^4,\quad
R_1(s)=\frac{25}{3}+7s-\frac{109}{3}s^2-22s^3. \label{16}
\end{equation}
The integral degree of circular polarization of the incoherent
radiation in a crystal we define by the ratio
$\xi_2^{inc}=I_{1}^{inc}/I_{0}^{inc}$.

The integral degree of circular polarization in the tungsten crystal
(axis $<111>$, the temperatures T=100 K)
$\xi_2^T=I_{1}(\varepsilon)/I(\varepsilon)$ Eq.(\ref{9}) is shown in
Fig.6 (the curve T), as well as the coherent degree
$\xi_2^{coh}=I_{1}^{coh}/I_{0}^{coh}$ Eq.(\ref{10}) (the curve 1)
and the incoherent degree $\xi_2^{inc}=I_{1}^{inc}/I_{0}^{inc}$
Eq.(\ref{16}) (the curve 2) as a function of incident electron
energy $\varepsilon$. In low energy region ($\varepsilon \leq 1$~GeV
the contribution of incoherent mechanism dominates (let us remind
that the intensities of the incoherent and coherent radiation become
equal at $\varepsilon \simeq 0.7$~GeV). At higher energies the
intensity $I_0^{coh}(\varepsilon)$ dominates while the intensity
$I_0^{inc}(\varepsilon)$ decreases monotonically \cite{BK1}.
Correspondingly the curve $\xi_2^{coh}$ tends to the curve
$\xi_2^T$. At extremely high energy $\varepsilon > 10^6$~GeV
$\xi_2^{coh}$ tends to the external field limit: $\xi_2^{coh}$=11/16
(see Eq.(4.88) in \cite{BKS}).

The next terms of decomposition of the total intensity
$I(\varepsilon)$ over $\nu_0^2$ describe the LPM effect in the
radiation process. The contribution of the LPM effect in the total
intensity of radiation $I$ Eq.(\ref{9}) is defined in \cite{BK1} as
$I^{LPM}=I_0 - I_0^{coh} -I_0^{inc}$. The relative contribution
(negative since the LPM effect suppresses the radiation process)
$\Delta=-I^{LPM}/I_0$  in the maximum is of the order of percent
(see Fig.3 in \cite{BK1}). Similarly we define the relative
influence of multiple scattering on the photon integral circular
polarization as
\begin{equation}
\Delta_{\xi}=-\frac{\xi_2^T-\xi_2^{ci}}{\xi_2^{ci}}=\Delta_1-\Delta,\quad
\xi_2^{ci}=\frac{I_1^{coh}+I_1^{inc}}{I_0^{coh}+I_0^{inc}},\quad
\Delta_1=\frac{I_1^{coh}+I_1^{inc}}{I_1}-1.
 \label{17}
\end{equation}
The function of $\Delta_{1}(\varepsilon)$ (per cent) is shown in
Fig.7, it attains the maximum $\Delta_{1} \simeq 2.0 \%$ at
$\varepsilon \simeq 1.4$~GeV, while $\Delta_{\xi}(\varepsilon)$
attains the maximum $\Delta_{\xi} \simeq 1.4 \%$ at $\varepsilon
\simeq 1.8$~GeV and $\Delta(\varepsilon) \simeq 0.9 \%$ at
$\varepsilon \simeq 0.3$~GeV \cite{BK1}. One can see that maxima of
corresponding functions are slightly shifted with respect each other
and the highest maximum has $\Delta_{1}$. From the other side, the
behavior of all functions $\Delta$, $\Delta_{1}$ and $\Delta_{\xi}$
are quite similar: just as in the total intensity of radiation the
suppression of integral polarization due to the multiple scattering
is concentrated in the interval of moderate energies $\varepsilon <
10~ $ GeV and scale of effect is of the order of percent.

\section{Conclusion}
\setcounter{equation}{0}

In this paper the spectrum of radiation from an electron of
intermediate energy (a few GeV for heavy elements) moving in
oriented crystal is calculated for the first time. The interplay of
the coherent and the incoherent parts is essential for formation of
the spectrum. Just in this situation the effects of multiple
scattering of projectile appear. The same is true also for depending
on polarization part of the spectral intensity.

In an oriented crystal at motion of an electron near the chain of
atoms (the axis) the atom density on the trajectory is much higher
than in an amorphous medium. As a result, the parameter
characterizing the influence of multiple scattering on the radiation
process in a medium in absence of an external field ($\nu_0^2 \sim
\varepsilon/\varepsilon_0$) becomes of the order of unity at enough
low energy (values of $\varepsilon_0$ for tungsten and germanium are
given in Table 1). From the other side, due to high density of atoms
at the trajectory near axis, the strong electric field of axis acts
on the electron. This action diminishes the radiation formation
length and expands the characteristic angles of photon emission and
hence weakens the influence of multiple scattering on the radiation
process. So, one has to use the general expression for the radiation
intensity which takes into account both the crystal effective field
(the coherent mechanism) and the multiple scattering (the incoherent
mechanism) for study of the characteristics of radiation. Such
expression for the unpolarized case $dI_0(\varepsilon, y)$ was
obtained in \cite{BK1} and the polarization term $dI_1(\varepsilon,
y)$ derived here (see Eq.(\ref{2})). The two first terms of
decomposition of $dI_{\xi}(\varepsilon, y)$ over the parameter
$\nu_0^2$ define the coherent and incoherent radiation. It should be
noted that in the incoherent contribution the influence of
crystalline field is taken into account. Other terms of the
decomposition represent influence of the crystalline field on the
multiple scattering (on the LPM effect).

Since in an amorphous medium the LPM effect for the whole spectrum
can be observed (for heavy elements) only in TeV energy range (see
e.g. \cite{BK2}) the possibility to study the influence of multiple
scattering on radiation process in GeV energy range is evidently of
great interest.

In the present paper the detailed analysis of the spectral and the
polarization properties of radiation is performed. The influence of
different mechanisms of photon emission on general picture of event
is elucidated. At high energy $\varepsilon \gg \varepsilon_0$  the
influence of the multiple scattering on the radiation intensity is
suppressed strongly (the coherent contribution dominates) and only
in the very end of the spectrum at $u \geq \varepsilon/\varepsilon_0
\gg 1~(1-\omega/\varepsilon \leq \varepsilon_0/\varepsilon \ll 1)$
the incoherent radiation  becomes essential. In this part of the
spectrum the nearly complete (with accuracy $\sim (\varepsilon_0/
\varepsilon)^2$) helicity transfer from electron to photon occurs
and the scaling defined by Eq.(\ref{a15}) takes place. For any
energy $\varepsilon$ the maximum value of the LPM effect is around 7
\% is situated at $u_m \sim
3\varepsilon/\varepsilon_0~(\omega_m=\varepsilon u_m/1+u_m)$. For
$\omega=\omega_m$ the incoherent contribution dominates, its
contribution is one order of magnitude higher than the coherent one.
So at $\varepsilon \gg \varepsilon_0$ the maximum of the LPM effect
is situated in very end of the spectrum where the mentioned scaling
holds. It should be noted the very right end of spectrum is
described by the Bethe-Maximon formulae with independent on the
electron energy crystal corrections (compare e.g. with Eq.(8) in
\cite{BK1}). For illustration of the discussed effect we considered
the low energy $\varepsilon \leq \varepsilon_0$, where both the
coherent and incoherent contributions are essential, while the
mentioned scaling is only approximate one. This energy region is
suitable for experimental study.

The polarization effects in radiation for the intermediate energy is
analyzed for the first time. It is shown that the influence of
multiple scattering on the depending on polarization part of
intensity spectrum $dI_1(\varepsilon, y)$ is very close to the LPM
effect in the unpolarized part $dI_0(\varepsilon, y)$.

\vspace{0.5cm}

{\bf Acknowledgments}

The authors are indebted to the Russian Foundation for Basic
Research supported in part this research by Grant 06-02-16226.

\newpage

{\bf Figure captions}

\vskip8mm

{\bf Fig.1} The spectral distribution of radiation in tungsten, axis
$<111>$, temperature T=100 K, with taking into account all
mechanisms of photon emission. The spectral inverse radiation length
(in cm$^{-1}$) $dI_0(\varepsilon,y)/\varepsilon dy$, see
Eq.(\ref{2}), is shown vs $y=\omega/\varepsilon$ for different
energies: curve 1 is for $\varepsilon=0.3$~GeV, curve 2 is for
$\varepsilon=1$~GeV and curve 3 is for $\varepsilon=3$~GeV.
\vskip2mm

{\bf Fig.2} The different contributions to the photon spectrum for
electron energy $\varepsilon=3$~GeV, axis $<111>$, temperature T=100
K. The curve 1 is $dI_0(\varepsilon, y)$, the curve 2 is the
coherent spectrum $dI_0^{coh}(\varepsilon, y)$, the curve 3 is the
incoherent spectrum $dI_0^{inc}(\varepsilon, y)$, the curve 4 is the
sum $dI_0^{coh}(\varepsilon, y)+dI_0^{inc}(\varepsilon, y)$, and the
curve 5 is the difference $dI_0^{coh}(\varepsilon, y) +
dI_0^{inc}(\varepsilon, y) - dI_0(\varepsilon, y)$.

\vskip2mm

 {\bf Fig.3} The LPM effect for spectral distribution of
radiation in tungsten, axis $<111>$, temperature T=100 K. The
function  $\Delta_s(y)$ Eq.(\ref{s15}) is shown vs
$y=\omega/\varepsilon$. Curve 1 is for $\varepsilon=0.3$~GeV, curve
2 is for $\varepsilon=1$~GeV and curve 3 is for $\varepsilon=3$~GeV.

{\bf Fig.4} the same as in Fig.2 but for T=293 K

\vskip2mm

{\bf Fig.5}  Degree of the circular polarization of radiation in
tungsten, axis $<111>$, temperature T=100 K. The value of
$\xi_{2}(y)$ Eq.(\ref{8}) is shown vs $y=\omega/\varepsilon$. The
curves for $\varepsilon=0.3, 1, 3$~GeV coincide.

\vskip2mm

{\bf Fig.6} The integral degree of the circular polarization in
tungsten, axis $<111>$, temperature T=100 K. The functions are shown
vs electron energy in GeV. Curve 1 is for the coherent radiation
($\xi_2^{coh}=I_1^{coh}(\varepsilon)/I_0^{coh}(\varepsilon)$), curve
2 is for the incoherent radiation
($\xi_2^{inc}=I_1^{inc}(\varepsilon)/I_0^{inc}(\varepsilon)$), curve
T is $\xi_2^T=I_{1}(\varepsilon)/I_0(\varepsilon)$ (see
Eq.(\ref{9})).

\vskip2mm

{\bf Fig.7}  Influence of multiple scattering on circular
polarization of emitted radiation described by the function
$\Delta_{1}(\varepsilon)$ Eq.(\ref{17}) in tungsten, axis $<111>$,
temperature T=100 K.

\newpage
\begin{table}
\begin{center}
{\sc Table 1}~ {Parameters of radiation process of the tungsten
crystal, axis $<111>$ and germanium crystal, axis  $<110>$ for two
temperatures T}
\end{center}
\begin{center}
\begin{tabular}{*{10}{|c}|}
\hline Crystal& T(K)&$V_0$(eV)&$x_0$&$\eta_1$&$\eta$&
$\varepsilon_0$(GeV)&$\varepsilon_t$(GeV)&$\varepsilon_s$(GeV)&$h$ \\
\hline W & 293&417&39.7&0.108&0.115&7.43&0.76&34.8&0.348\\
\hline W &100&355&35.7&0.0401&0.0313&3.06&0.35&43.1&0.612\\
\hline Ge & 293 & 110& 15.5
&0.125&0.119&148&1.29&210&0.235\\
\hline Ge & 100 & 114.5& 19.8
&0.064&0.0633&59&0.85&179&0.459\\
\hline
\end{tabular}
\end{center}
\end{table}


\newpage
\begin{thebibliography}{99}
\bibitem{BK1} V. N. Baier, and  V. M. Katkov,
Phys. Lett.,A {\bf 353} (2006) 91.
\bibitem{KMU} K. Kirsenom {\it et al}, Phys. Rev. Lett.,
{\bf 87} (2001) 054801.
\bibitem{BKi} A. Baurichter {\it et al}, Phys. Rev. Lett.,
{\bf 79}  (1997) 3415.
\bibitem{BK}  V. N. Baier, V. M. Katkov,
 Nucl.Instr.and Meth B, {\bf 234} (2005) 106.
\bibitem{BKS} V. N. Baier, V. M. Katkov and V. M. Strakhovenko,
{\em Electromagnetic Processes at High Energies in Oriented Single
Crystals}, World Scientific Publishing Co, Singapore, 1998.
\bibitem{BK3} V. N. Baier and V. M. Katkov,
Phys.Rev. D {\bf 62} (2000) 036008.
\bibitem{BK2} V. N. Baier and V. M. Katkov,
Phys.Rep. {\bf 409} (2005) 261.


\end{thebibliography}
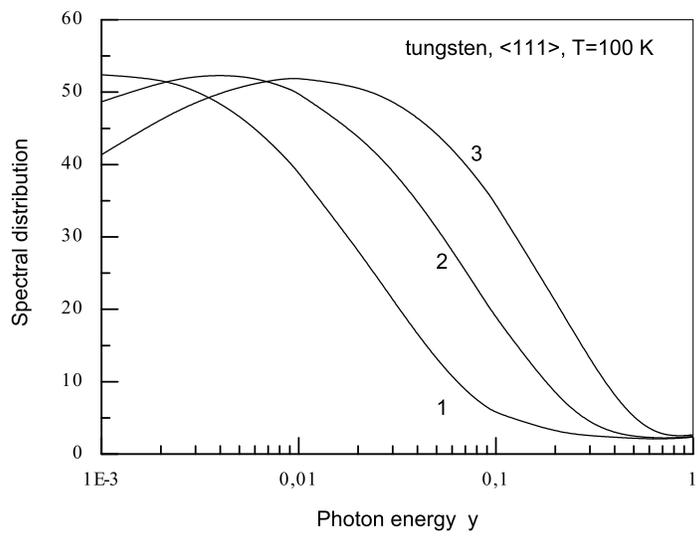
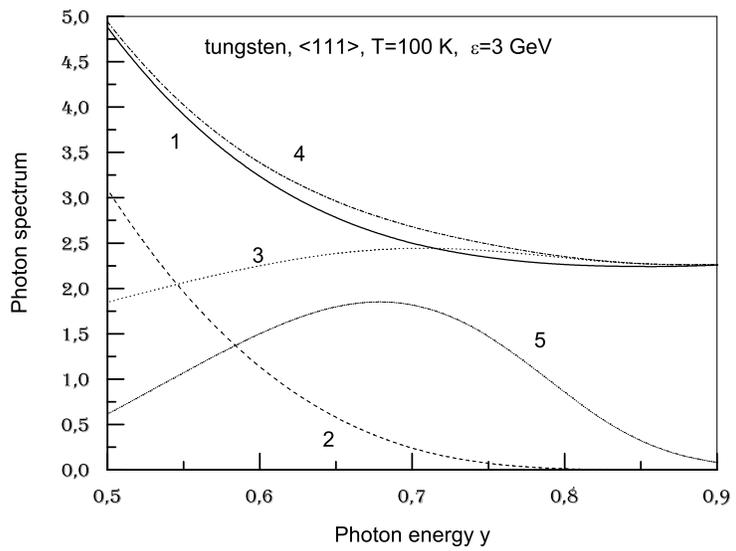
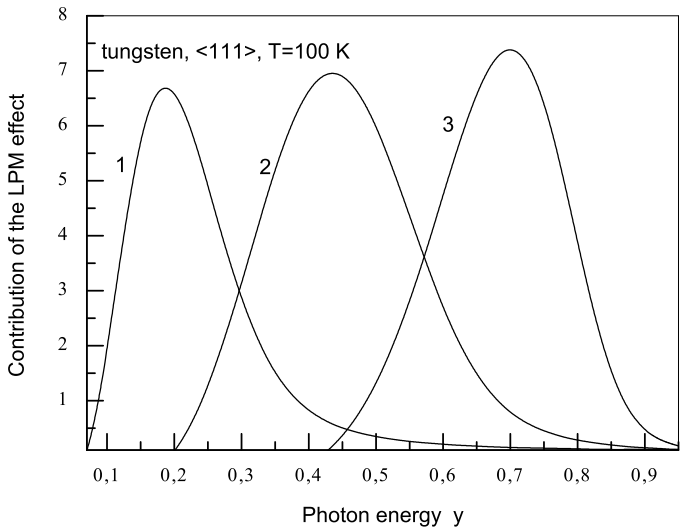
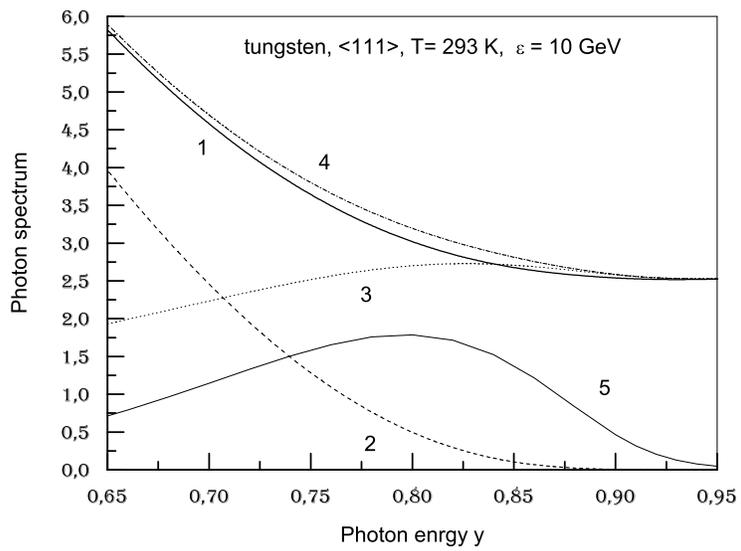
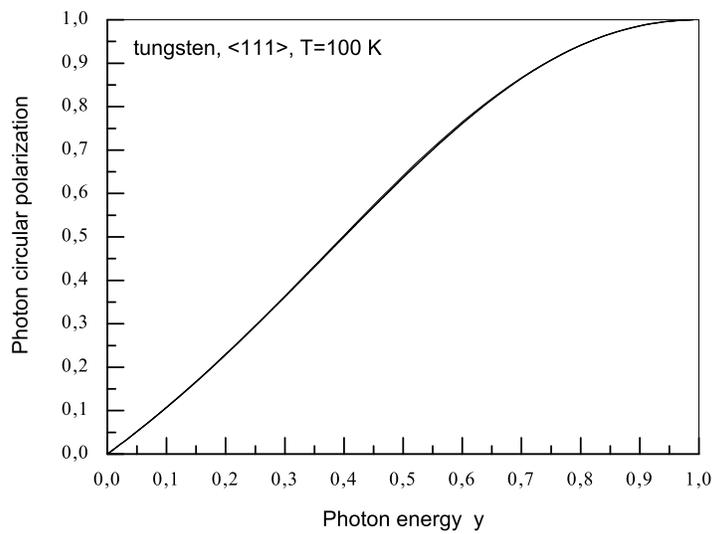
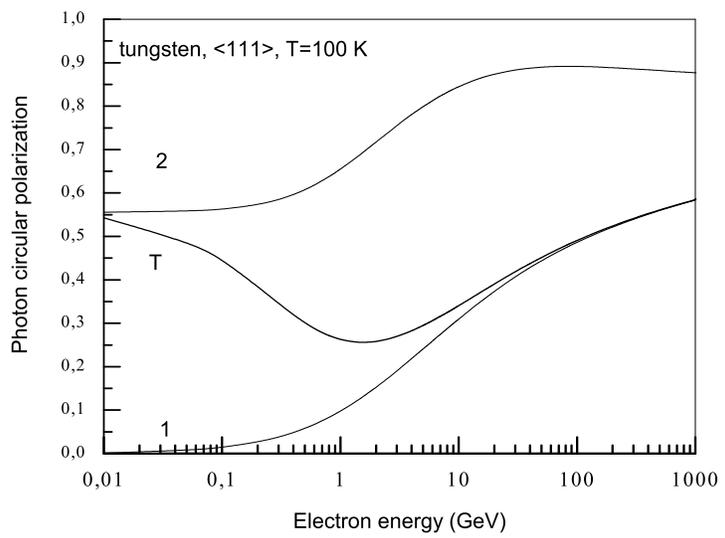
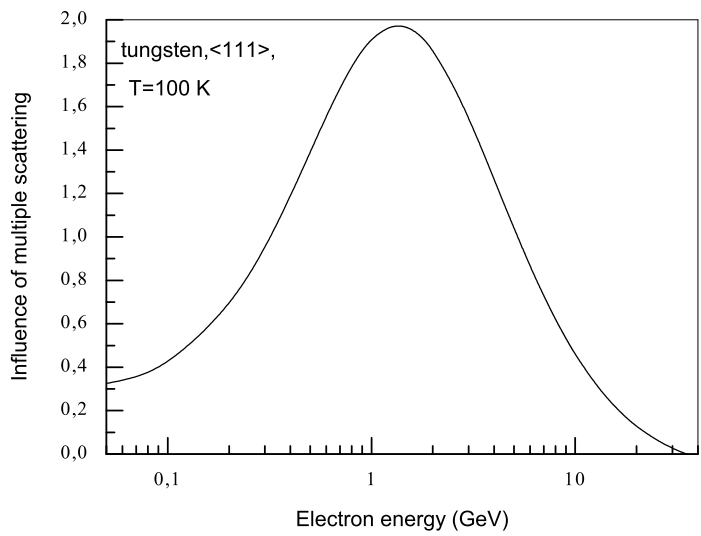
\end{document}